\documentclass[12pt]{article}
\usepackage{latexsym}
\newcommand{\f}{\begin{equation}}
\newcommand{\ff}{\end{equation}}
\newcommand{\fa}{\begin{eqnarray}}
\newcommand{\ffa}{\end{eqnarray}}
\newcommand{\G}{\Gamma}
\title{ Eleven dimensional supergravity as a constrained
topological field theory}
\author{Yi Ling$^{\dag}$ and Lee Smolin$^{*}$
\thanks{email:$\dag$ ling@phys.psu.edu, $^{*}$ smolin@phys.psu.edu}\\
\centerline{\it Center for Gravitational Physics and Geometry}\\
\centerline{\it Department of Physics}\\ \centerline {\it The
Pennsylvania State University}\\ \centerline{\it University Park,
PA, USA 16802} \\ \centerline{and} \\ \centerline{\it The Blackett
Laboratory}  \\ \centerline{\it Imperial College of Science,
Technology and Medicine}\\ \centerline{\it South Kensington,
London SW7 2BZ, UK}}
\begin{document}
\maketitle
\begin{abstract}
\baselineskip=20pt We describe a new first-order formulation of
$D=11$ supergravity which shows that that theory can be understood
to arise from a certain topological field theory by the imposition
of a set of local constraints on the fields, plus a lagrange
multiplier term. The topological field theory is of interest as
the algebra of its constraints realizes the $D=11$ supersymmetry
algebra with central charges.
\end{abstract}
\vfill \eject
\section{Introduction}
\baselineskip=20pt

Eleven dimensional supergravity has been an object of fascination
since it was first discovered more than twenty years
ago\cite{CJS}.  It is the highest dimension in which a
supergravity theory exists, and it  contains all of the lower
dimensional supergravities as dimensional reductions.   It has
also been proposed as a description of the classical limit of a
phase of $\cal M$ theory\cite{mtheory}, which is not itself
described by any consistent string theory.  Anyone wanting to
understand the dynamical structure of the different
supergravities, and their inter-relations with each other under
reductions and various kinds of duality transformations are
recommended to start with the $11$ dimensional theory.

It is thus very interesting to wonder whether $11$ dimensional
supergravity might be made into a quantum theory directly, using
some non-perturbative approach. It is true that it is
non-renormalizable in perturbation theory\cite{nonren}, but so are
general relativity and the $N=1,2$ supergravities in four
dimensions, and this has not prevented a great deal of progress
being made understanding the exact structure of their canonical
and path integral quantum theories\cite{lp1}-\cite{stringsfrom}. A
beautiful and robust structure was found that could not have been
seen in  perturbation theory, which leads to definite physical
predictions such as the discrete spectra of the area and volume
operators\cite{spain,sn1,sn2}. The eigenstates of these operators
provide a basis of states for background independent quantum
theories of gravity, which are the spin network
basis\cite{sn1,sn2}. These are constructed in terms of simple
combinatorial and representation theory data, are closely related
to the fusion algebras of conformal field
theory\cite{qdeform,tubes,mpaper} and exist also for
supersymmetric theories\cite{supersn,super}.

It is also clear now that these theories have rigorously defined
hamiltonian\cite{rigorous} and path
integral\cite{foam,barrettcrane,evolving} formulations and so do
exist as examples of diffeomorphism invariant quantum field
theories. This existence is independent of the question of whether
they have classical limits which reproduce classical general
relativity or supergravity. This is presently an open question for
general relativity and supergravity in $d=4$\cite{trouble}.   Even
if the particular dynamics of quantum general relativity in $d=4$
leads to a theory which is sensible at the Planck scale but lacks
a good classical limit, it is still reasonable to conjecture that
the robust and theory independent features of the background
independent formulation of quantum gravity uncovered in loop
quantum gravity will play a role in a background independent
formulation of $\cal M$ theory\cite{tubes,mpaper,cubicmatrix}.

There is a simple reason to expect that the classical limit will
exist for a background independent of $11$ dimensional
supergravity, but not for $d=3+1$ general relativity, which is
that whenever such a limit exists one expects a sensible
perturbation theory to exist around any classical background that
arises in a classical limit. There is no such consistent
perturbation theory for pure four dimensional general relativity
but there is one for many, if not all, consistent
compactifications of $11$ dimensional supergravity-these are of
course exactly the perturbative string theories. In fact, there is
a simple argument that where such a limit exists the weakly
coupled excitations will be described by a string
theory\cite{stringsfrom}.

It is then sensible to suppose that one way to uncover the
structure of the states and operators that will go into the
background independent formulation of $\cal M$ theory is to make
an exact canonical quantization of $11$ dimensional supergravity
and discover what structures play the role of the spin networks
and super-spin networks in $D=3+1$. The first step in any such
attempt must be to have a suitable form for the action.  What is
required to make progress in a non-perturbative approach to
quantization is to have a first order polynomial form of the
action, which has the property that it arises from a topological
field theory by the imposition of a set of constraints.  The
reason is it is by now understood that all the beautiful results
which have followed from the use of the Ashtekar-Sen and related
variables are due ultimately to the fact that in $4$ dimensions
general relativity and supergravity (at least up to
$N=2$\cite{super}) can be understood as arising in this way from
topological field theories\footnote{Indeed, such a form of the
theory was understood earlier, by Plebanski\cite{plebanski}.  The
significance of this form was only realized after Sen\cite{sen}
discovered the equivalent Hamiltonian variables, in an attempt to
understand the canonical structure of supergravity. This was then
formalized by Ashtekar\cite{ashtekar} for the canonical theory and
in \cite{tedlee,sam} for the lagrangian theory. Attempts to relate
this form to topological field theory then led to a rediscovery of
Plebanski's action\cite{CDJ} and its extension to supergravity.}.
A further reason for taking this route is that in the presence of
appropriate boundary conditions it leads to holographic
formulations of quantum general relativity and supergravity in
which the Bekenstein bound is realized naturally because the
boundary theories are built from the state spaces of Chern-Simons
theory\cite{linking,hologr,superholo}\footnote{This method has
also been used recently to construct an explicit description of
the quantum geometry of the black hole horizon\cite{bhhorizon}.}.

The main goal of this paper is then to present a formulation of
the $11$ dimensional supergravity as a constrained topological
field theory.  In the next section we introduce an $11$
dimensional topological field theory which has the property that
its algebra of first class constraints, which generate the gauge
transformations of the theory, reproduces exactly the $11$
dimensional super-Poincare algebra, including its central charges.
The theory is introduced in the next section and the canonical
analysis is given in section 4.  In section 3 we show how the
eleven dimensional supergravity action arises by imposing a
certain set of constraints on the topological field theory. We
then arrive at the supergravity action in the form given by
Fre\cite{Fre2} and d'Auria and Fre\cite{dAF}. A feature of that
form of the theory, which plays an important role as well in our
formulation, is the presence of two abelian potentials, which are
a six form and a three form. Although we are not completely
certain of the correct way to express it, we believe it likely
that the formulation of a gravitational theory as a constrained
topological field theory is closely related to the idea of
formulating it in terms of free differential algebras, which was
pursued by Fre and collaborators\cite{fda}\footnote{Another
approach to the relationship between topological field theory and
supergravity is described in \cite{Chamseddine1,Chamseddine2}.}.

This paper represents only the first step of a program of
quantization of $11$ dimensional supergravity. Still to be
investigated is the implications for the canonical structure of
the full $11$ dimensional supergravity action and the possible
existence of a quantization using the methods of loop quantum
gravity.

In the next section we introduce the topological quantum field
theory in $11$ dimensions and in section 4 we derive its canonical
formalism and compute its constraint algebra. We find that it is
first class and that the algebra of constraints does reproduce
weakly\footnote{That is up to the constraints that say that the
curvatures vanish in a topological field theory} the
super-Poincare algebra in $11$ dimensions with central charges.
Along the way, in section 3, we show how constraints and lagrange
multiplier terms may be added to the topological field theory to
arrive at the full $11$ dimensional supergravity action, in the
form given by Fre\cite{Fre2}.

\section{TQFT for the $11$ dimensional super-Poincare algebra }

The first step in our construction is to find the TQFT whose
algebra of constraints reproduces the supersymmetry algebra of
supergravity in $11$ dimensions, including the central extensions.
This algebra has the form\footnote{We list the convention used in
this paper. $(i)$ Indices $\mu, \nu,...=0, 1, ..., 10$ are
spacetime indices while $i, j, ...=1, 2, ..., 10$ are used for
spatial indices; $(ii)$ The capital $ A, B,...=1, 2, ..., 32$
stand for the $Sp(32)$ spinor indices; $(iii)$  $a, b, ..., =0, 1,
...,10$ represent $SO(10,1)$ indices. We sometimes use a condensed
index notation in which $a_{p}=a_{1}a_{2}\ldots a_{p}$.}

\f \left\{Q^A,Q^B\right\}=\Gamma_a^{AB}G^a+\Gamma_{ab}^{AB}G^{ab}+
\Gamma_{a_5}^{AB}G^{a_5}, \label{alg} \ff where \f
\Gamma_{a_p}^{AB}:=\Gamma_{[a_1}\Gamma_{a_2}...\Gamma_{a_p]}^{AB}.
\ff

Here $G^{a}$ must be the generator of spacetime diffeomorphisms,
so that $G^{0}$ must be related to the Hamiltonian and $G^{i}$ to
the diffeomorphism constraints of the theory, while $G^{ab}$ and
$G^{a_5}$ are the central extensions. This can be understood to be
a contraction of $Osp(1|32)$\cite{dAF}.

One place to begin is with the fields of a gauge theory for the
superalgebra $Osp(1|32)$ in $10+1$ dimensions.  The generators of
the superalgebra consist of the translation generator $G_a$,
Lorentz generator $J_{ab}$, supersymmetry generators $Q^A$ and
five-index antisymmetric generator $G^{[abcde]}\equiv G^{a_5}$.
One starting point would be to define a 1-form superconnection
associated with those generators: \f \cal
A\mit_{\mu}:=A^{ab}_{\mu}J_{ab}+e^a_{\mu}G_{a}+\Psi^A_{\mu}Q_A+
A^{a_5}_{\mu}G_{a_5}, \ff where $\Psi^{A}$ are $32$ component
Majorana spinors. We may  then introduce a $BF$ action as \f \cal
I\mit=\int_{M}dx^{11}\cal B\mit\wedge\cal F\mit,\label{BFOSP} \ff
where $\cal B\mit$ is a super nine form and $F$ is the curvature
of $Osp(1|32)$ defined by \f \cal F\mit= d\cal A\mit+\cal
A\mit\wedge\cal A\mit . \ff

Unfortunately this route  seems not to  lead to the standard $11$
dimensional supergravity\footnote{But it is of interest and has
been pursued in \cite{d=11cs,horava}.}.  The difference seems to
be that the relevant gauge group for 11 dimensional supergravity,
at least at the classical level, is the Super-poincare group,
which is a contraction of $Osp(1|32)$.  As a result, the central
generators are realized in supergravity by functionals of a three
form abelian gauge field rather than by a standard component of a
connection one-form.  The basic mystery of the construction of
$11$ dimensional supergravity (as well as many of its lower
dimensional reductions) is   how such a field may be seen to arise
from a gauge theoretic structure as does the connection of
spacetime.  From the point of view in which gravitational theories
are understood as constrained topological field theories, this
mystery can be solved, for topological field theories can indeed
be constructed based on gauge theories of Abelian $p$-forms and
they can be quantized using the methods of loop quantum
gravity\cite{prodolfo,diffinv,lee11cs}. Indeed as we shall see
here, one can construct a topological field theory whose
constraint algebra realizes perfectly the full super-Poincare
algebra, with the abelian $p$ forms realizing the central
charges\footnote{Thus, while we solve the problem of encoding the
central charges in an algebra of canonical constraints, in a way
that leads under suitable constraints to $11$ dimensional
supergravity, we do not solve the problem of what all this may
have to do with $Osp(1|32)$.}.

In order to achieve this we have found it necessary to introduce
not only the three form gauge field $a_{\mu \nu \rho}$ but its
dual, which is a six form field $b_{\alpha \beta \gamma \delta
\epsilon \phi}$.  This leads us to a formalism which is similar to
that of D'Auria and Fre\cite{Fre2,dAF}. Indeed, as our results
show, there is likely a close relationship between their
conception of supergravity based on Cartan integrable algebras and
the more recent conception of a gravitational theory as a
constrained topological field theory.

We now introduce our $11$ dimensional super-Poincare topological
field theory.

To begin with we define the topological field theory associated to
the $11$ dimensional super-Poincare algebra, unextended by central
charges.  The gauge field is then of the form, \f {\cal
A}_{\mu}:=A^{ab}_{\mu}J_{ab}+e^a_{\mu}G_{a}+\Psi^A_{\mu}Q_A. \ff
The components of the curvature are given by \f
F^{ab}=dA^{ab}-A^{ac}\wedge A_c^b, \ff \f F^a=de^a-A^{ab}\wedge
e_b-{i\over 2}\Psi_A\wedge {\Gamma^{aA}}_B\Psi^{B}, \ff \f
F^A=D\Psi^A=d\Psi^A-{1\over 4}A^{ab}\Gamma_{abB}^{\ A} \wedge
\Psi^B. \ff To represent the central charges we introduce the
three form $a_{\mu_3}$ and its dual which is a 6-form field
$b_{\mu_6}$. We introduce their curvatures, \f F^\otimes=db-15
da\wedge a- {\imath \over 2} \rho^5 ,\ff \f F^{\Box}=da- {1\over
2} \rho^2, \ff where \f \rho^p_{a_p}:={\Psi}_A\wedge
{\Gamma^{a_pA}}_{B}\Psi^B \wedge E^p_{a_p}, \ff and \f
E^p_{a_p}=e_{a_1}\wedge e_{a_2}...\wedge e_{a_p}. \ff Using these
curvatures we then write a TQFT, \f I^{TQFT}={ - 1\over g^{2 } }
\int B_{ab}\wedge F^{ab}+B_{a}\wedge F^{a }+ B_A\wedge F^A
+B^\Box\wedge F^{\Box} +B^\otimes \wedge F^{\otimes}. \label{TQFT}
\ff

The $B$'s are lagrange multipliers which have the form degree
indicated.  The field equations are \f
F^{ab}=F^{a}=F^{A}=F^{\Box}=F^{\otimes}=0, \ff and \f {\cal
D}\wedge B^{ab}- e^{[a} \wedge B^{b]}- {1\over 4}
\Psi_{A}{\Gamma^{abA}}_{B}\wedge B^{B}=0, \ff \f {\cal D}\wedge
B^{a}-\Psi_{A}{\Gamma^{abA}}_{B}\Psi^B\wedge e^b\wedge
B^{\Box}-{5i \over 2}\Psi_{A}{\Gamma^{ab_4A}}_{B}\Psi^B\wedge
e_{b_4}\wedge B^{\otimes} =0,\ff \f {\cal D}\wedge B^{A}-
i\Gamma^A_{aB}\Psi^{B}\wedge B^{a}-\Gamma^A_{abB}\Psi^{B}\wedge
E^{ab}\wedge B^{\Box}- i\Gamma^A_{a_5B}\Psi^{B}\wedge
E^{a_5}\wedge B^{\otimes}=0, \ff \f d\wedge B^{\otimes}=0, \ff \f
d\wedge B^{\Box}-30da\wedge B^{\otimes}+15a\wedge d\wedge
B^{\otimes}=0. \ff

In section (4) we will discuss the canonical formulation of the
action (\ref{TQFT}) and show that its algebra of first class
constraints replicates the superalgebra (\ref{alg}).

\section{Constraining the $TQFT$ to get supergravity in
$11$ dimensions}

We obtain an action for $11$ dimensional supergravity by adding
constraint and lagrange multiplier terms, \f
I^{SUGRA}_{11D}=I^{TQFT}+I^{CONST.}+I^{F_4}, \ff where
\begin{eqnarray}
I^{CONST.}& =& \int \lambda_{ab}\wedge (B^{ab}-{1 \over 9l^9}
E^{*ab})+\lambda_{A}\wedge (B^A-{2 \over l^8}
\Gamma^{A}_{a_8B}\Psi^B\wedge E^{a_8}) \nonumber\\&&
+\lambda_{a}\wedge (B^{a}-{7\imath \over 30}
E_{7}^{ab_{6}}\Psi_{A}\Gamma^{c_5A}_{B}\Psi^{B}\epsilon_{b_{6}c_{5}})
\nonumber \\&& + \lambda_{\otimes}\wedge (B^{\otimes}-56da)+
\lambda_{\Box}(B^{\Box}+56\imath \rho^{5}),
\end{eqnarray}
and \f I^{F_4}=\int -2 F_{a_4}R^{\Box}\wedge
E^7_{b_7}\epsilon^{a_4b_7}+ {1\over 330} F_{a_4}F^{a_4}E_{11}^{*}.
\ff It is not difficult to see that the variations of the lagrange
multipliers $\lambda$ reproduce the $D=11$ supergravity action in
the form given by Fre\cite{Fre2}. \fa I^{SG}&=&{ - 1\over g^{2 } }
\int {1 \over 9l^9} E^{*ab}\wedge F^{ab}+ {7\imath \over 30}
E_{7}^{ab_{6}}\Psi_{A}\Gamma^{c_5A}_{B}\Psi^{B}\epsilon_{b_{6}c_{5}}
\wedge F^{a }+ {2 \over l^8} \Gamma^{A}_{a_8B}\Psi^B\wedge
E^{a_8}\wedge F^A \nonumber \\ &&+ 56\imath \rho^{5}\wedge
F^{\Box} + 56da \wedge F^{\otimes} \nonumber \\ &&-2
F_{a_4}R^{\Box}\wedge E^7_{b_7}\epsilon^{a_4b_7}+ {1\over 330}
F_{a_4}F^{a_4}E_{11}^{*}. \ffa Elimination of the lagrange
multipliers then leads to the theory in the original
form\cite{CJS}.

We not that the lagrange multiplier terms $I^{F_4}$ are necessary
to get the $da^{2}$ terms in the supergravity action.  Were they
absent the $a$ field would have dynamics only from the
Chern-Simons like terms $a\wedge da \wedge da$.  One interesting
question that this approach should be able to answer is why
supersymmetry requires both the Maxwell and Chern-Simons like
terms in the supergravity action.

\section{Canonical formulation of the $11D$ TQFT}

We now describe to the canonical decomposition of the TQFT given
by (\ref{TQFT}).  Our main goal here is to find the algebra of its
constraints.  We assume that the eleven dimensional spacetime
${\cal M}^{11}$ has the form ${\cal M}^{11}=\Sigma^{10}\times R$
where $\Sigma^{10}$ is a compact ten dimensional manifold.  We
then make a $10+1$ decomposition to find that\footnote{Note that
the decomposition is much simpler than in the standard case as
there is no metric and hence no lapse and shift.  The
decomposition is purely a matter of pulling back forms.} \fa
I^{TQFT}& = & \int dt\int_{\Sigma^{10}} d^{10}x  \left \{{1\over
2}  \pi^{iab}\dot{A}_{iab}+\pi^{ia}
\dot{e}_{ia}+\pi^{iA}\dot{\Psi}_{iA}+ {1\over 6!}
p^{i_6}\dot{b}_{i_6}+{1\over 3!} R^{i_3}\dot{a}_{i_3} \right.
\nonumber\\ && +B_0^{ab}\wedge F_{ab}+B_0^A \wedge
F_A+B_0^{a}\wedge F_{a}+B_{0}^{\Box}\wedge F^{\Box}+
B_{0}^{\otimes}\wedge F^{\otimes} \nonumber \\  && \left.
+A_{0ab}J^{ab}+\Psi_{0A}Q^A+e_{0a}G^a+a_{0ij}
G_2^{ij}+b_{0i_{5}}G_{5}^{i_{5}} \right \}, \ffa

where now all forms are in the $10$ dimensional space.

The expressions for the canonical momenta are,
\begin{eqnarray}
    \pi^{i}_{ab}& = & {-2 \over g^{2 } } (B_{ab}^{*})^{i},  \nonumber \\
    \pi^{i}_{a}& =& { 1\over g^{2 } }(B_{a}^{*})^{i},  \nonumber \\
    \pi^{i}_{A}& =& { 1\over g^{2 } }(B_{A}^{*})^{i},  \nonumber \\
    p^{i_6}& =&{ 1\over g^{2 } } (B^{*}_{\otimes})^{i_6},  \nonumber \\
    R^{i_3} & =&  { 1\over g^{2 } }
    (B_{\Box}^{*})^{i_{3}}-{15 \over 3!} p^{i_{3}j_{3}}a_{j_{3}}.
\end{eqnarray}

The constraints take the form \fa Q^A &=& D_i\pi^{iA} -{\imath }
\pi^{ia}\Gamma^{A}_{aB}\Psi^B_{i} -\Psi^{B}_{i}\left \{
 \Gamma_{abB}^A ( R^{ijk}+{15\over 3!} p^{ijkl_{3}}a_{l_{3}}
)E^{ab}_{jk} -{\imath } \Gamma^A_{a_{5}B}
p^{ik_{5}}E_{k_{5}}^{b_{5}} \right \}, \\ G^a& =& D_k\pi^{ka}
-\Psi^{A}_{i}\Psi^{B}_{j} \left \{ \Gamma^{a}_{bAB} (
R^{ijk}+{15\over 3!} p^{ijkl_{3}}a_{l_{3}} ) e_{k}^{b} + {5 \imath
\over 2} \Gamma^{a}_{b_{4}AB}p^{ijk_{4}}E_{k_{4}}^{b_{4}} \right
\},\\ J^{ab}&=&{1\over 2} D_i\pi^{iab}-\pi^{i[a}e^{b]}_i-{1 \over
4} {\Psi}_{iA}{\Gamma^{abA}}_{B}\pi^{Bi},
\\G^{i_5}_{5}&=&{-1\over 5!}
\partial_i p^{ij_5}, \label{C5} \\ G_2^{ij}&=& {1\over 2} [\partial_k
R^{ijk}-{15\over 3!}(\partial_ka_{l_3})p^{ijkl_3}]. \label{C2}
\ffa

It's now straightforward to check the Poisson brackets of two
supersymmetric constraint functional satisfies weakly the relation
(\ref{alg}). More precisely, we find, \fa \{ Q^{A}(x),Q^{B}(y)
\}_{+} &= & \delta^{10}(x,y)  \left \{ \Gamma^{AB}_{a} G^{a}
\right.  \nonumber \\
 &&  - \Gamma^{AB}_{ab} \left [
+E_{jk}^{ab}(G_{2}^{jk} + {15 \over 3!} G_{5}^{jkl_{3}}a_{l_{3}} )
+{15 \over 3!}p^{ijkl_{3}}(F_{il_{3}}^{\Box}E_{jk}^{ab}+
a_{l_{3}}F_{ij}^{a}e_{k}^{b}) \right ]   \nonumber \\ && \left.
{-\imath }  \Gamma^{AB}_{ab^{4}}\left [
E_{k_{5}}^{ab_{4}}G_{5}^{k_{5}} +
p^{ikl_{4}}E_{l_{4}}^{b_{4}}F_{ik}^{a} \right ] \right \}. \label{SPB}\ffa It
is interesting to see that the constraints by which the curvatures
vanish are needed to close the algebra. It is also interesting
that in order to realize the central charge proportional to
$\Gamma^{AB}_{abcde}$ it is necessary to have the canonical
momenta and the constraint associated with the six form field $b$.

Similarly, we find that \fa \{ G^{a}(x),G^{b}(y) \} &= &
\delta^{10}(x,y) \left  \{ \Gamma^{ab}_{AB}\left [
2F_{ki}^{A}\Psi_{j}^{B}[R^{ijk} + {15\over 3!}
p^{ijkl_{3}}a_{l_{3}}] + \Psi^{A}_{i}\Psi^{B}_{j}( {1\over
4}G_{2}^{ij}+G_{5}^{ijl_{3}}a_{l_{3}}
+p^{ijl_{4}}F_{l_{4}}^{\Box}) \right ] \right. \nonumber \\
 &&+ \left. \Gamma^{ab}_{defAB}\left [{2\imath }F_{ki}^{A}\Psi_{j}^{B}
p^{ijkl_{3}}E_{l_{3}}^{def} +{\imath \over 4}
G_{5}^{ijl_{3}}E_{l_{3}}^{def} +{\imath }
p^{ijklm_{2}}F_{kl^{d}}E_{m_{2}}^{ef} \right ] \right \}.
\label{GwG} \ffa In both of these relations there are delicate
cancellations involving two and four fermion terms.  These involve
careful application of the Fiertz identities for $11$ dimensions.
It is also interesting that the central charges come into the
commutator of the translation generators (\ref{GwG}).  This may be
a clue as to how the whole structure may descend from some
$Osp(1|32)$ invariant framework.

We also find that, \f \{Q^{A}(x),G_{2}^{{ij}}(y)\} = {5\over 4}
\delta^{10} \Gamma^A_{abB} \left ( -G_5^{lmnij}\Psi_l^B
E_{mn}^{ab} +p^{lmnkij}[F_{kl}^B E_{mn}^{ab} +2F_{kl}^a \Psi_l^B
e_n^b ] \right ), \ff \f \{G^a(x),G_{2}^{{ij}}(y)\} = {5\over 4 }
\delta^{10} \Gamma^a_{bAB} \left ( -G_5^{lmnij}\Psi_l^A \Psi_m^B
e_n^b +p^{lmnkij}[ 2F_{kl}^A \Psi_m^B e_n^b + \Psi_l^A \Psi_m^B
F_{kn}^b ] \right ), \ff \f \{Q^{A}(x),G_{5}^{{ijklm}}(y)\}=
\{G^a(x),G_{5}^{{ijklm}}(y)\}=0. \ff

Thus, $G_2^{ij}$ and $G_5^{ijklm}$ form a supersymmetry multiplet.
We find also \f
 \{Q^{A}(x),G^{a}(y)\}=0.
\ff

The  commutators of the $J^{ab}$ are defined by the transformation
properties under the local lorentz group. \f \{J^{ab},J_{cd}\}=
\delta^{10}(x,y)  \delta^{[a}_{[c}J^{b]}_{d]}, \ff \f
\{J^{ab}(x),G^{c}(y) \}= \delta^{10}(x,y) \delta^{[a}_{c}G^{b]},
\ff \f \{J^{ab}(x) ,Q_{A}(y) \}= \delta^{10}(x,y)
\Gamma^{abB}_{A}Q_{B}, \ff \f \{J^{ab}(x) ,G_{2}^{ij}(y) \}=0, \ff
\f \{J^{ab}(x) ,G_{5}^{ijklm}\} =0. \ff Finally, we find that all
the other communtators vanish. Thus the constraint algebra does in
fact reproduce the super-Poincare algebra with central charges.

\section{Conclusions}

We have reported here the first step of a program to construct a
non-perturbative formulation of $\cal M$ theory by making a
background independent quantization of $11$ dimensional
supergravity.  The next step is to construct the quantization of
the $11$ dimensional topological quantum field theory, using the
methods of loop quantum gravity. This can be done both canonically
and through a path integral quantization using an extension of the
methods of spin foam\cite{foam} or evolving spin
networks\cite{evolving} to $p$-form gauge fields. Some work in
this direction already exists\cite{prodolfo,diffinv,lee11cs} and
this part of the program should go through directly. It is clear
from the form of the theory that this will involve extended
objects whose spacetime dimensions are $2,3$ and $6$.  These will
then give background independent objects corresponding to strings,
membranes and five-branes.

In the topological quantum field theory these will have trivial
dynamics and the states will be functionals only of homotopy
classes of the ten dimensional spacial manifold (which is fixed in
a canonical quantization). The problem will then be to reduce the
gauge invariance of the topological quantum field theory so as to
give rise to local degrees of freedom.  There are two ways to
accomplish this. The conservative, straightforward path will be to
construct the canonical quantization of the eleven dimensional
supergravity, by imposing the constraints in the above action
classically. This will involve a lot of tedious calculation, to
check the resulting algebra of constraints, but should be
nonetheless straightforward.  One will then impose the quantum
constraints rather than the vanishing curvature conditions on the
Hilbert space of states.

A second  route to the theory will be to follow Barrett and
Crane\cite{barrettcrane} and impose the constraints directly in a
path integral expression for the topological quantum field theory.

While these will involve a great deal of work, the key point is
that at every stage one will be working with a background
independent definition of the Hilbert space of the theory, whose
degrees of freedom have the correct dimensionality and
supersymmetry transformation properties to lead to strings,
membranes and fivebranes in the classical limit. It is difficult
to believe that something of value for the understanding of $\cal
M$ theory will not come out of such an investigation.

\section*{ACKNOWLEDGEMENT}

We are grateful to Clifford Johnson, Hermann Nicolai, Mike
Reisenberger and Kelle Stelle for discussions during the course of
this investigation as well as to the theoretical physics group at
Imperial College for hospitality during the course of this work.
This work was supported by the NSF through grant PHY95-14240 and a
gift from the Jesse Phillips Foundation.

\section*{Appendix A: Conventions and notations}
Here we give the conventions and notations in this paper.
We adopt the convention that if $\omega$ is a p-form, then:
\f \omega_p:=\omega_{\mu_1...\mu_p}dx^{\mu_1}\wedge
dx^{\mu_2}...\wedge dx^{\mu_p},\ff
or we could write it with abstract indices as,
\f\omega_{a_1a_2...a_p}=p!\omega_{\mu_1...\mu_p}(dx^{\mu_1})_{[a_1}(
dx^{\mu_2})_{a_2}...(dx^{\mu_p})_{a_p]},\ff
Where the antisymmetrization symbol is defined by,
\f [a_1...a_n]={1\over
n!}\sum_p(-)^{\delta_p}a_{p(1)}...a_{p(n)},\ff
where $\sum_p$ is the sum over permutations and $\delta_p$ is the
parity of the permutation.
Given any two forms such that one is p-form and the other one is q-form, we can
define the wedge product of these two forms as,
\f \omega_p\wedge\omega^{'}_q={11!\over
p!q!}\omega_{[a_1...a_p}\omega^{'}_{b_1...b_q]}.\ff
It's straightforward to show the wedge product has the following
property,
\f\omega_p\wedge\omega^{'}_q=(-1)^{p*q}\omega^{'}_q\wedge\omega_p.\ff
The exterior differential $d$ is a map from vector space of p-form
to that of $(p+1)$-form,
\f d\omega_p=(d\omega)_{ab_1...b_p}=(p+1)\nabla_{[a}\omega_{b_1...b_p]}.\ff
we can also show that
\f d(\omega_p\wedge \omega^{'}_q)=(d\omega_p)\wedge
\omega^{'}_q+(-1)^p\omega_p\wedge d(\omega^{'}_q).\ff
If $\omega_{11}$ is a 11-form on the manifold $\cal M$, we define
the integral of the form on the manifold as,
\f \int\omega_{11}={1\over
11!}\int\epsilon^{\mu_1...\mu_{11}}\omega_{\mu_1...\mu_{11}}d^{11}x,\ff
where $\epsilon^{\mu_1...\mu_{11}}$ is the volume element on $\cal
M$ such that
\f \epsilon^{\mu_1...\mu_{11}}\epsilon_{\mu_1...\mu_{11}}=11!,\ff
and locally if the manifold splits into space $\Sigma_{10}$ and
time $R$ which is denoted by the coordinate $0$, then an induced volume on
$\Sigma_{10}$ is given by,
\f\epsilon^{i_1...i_{10}}=\epsilon^{0\mu_2...\mu_{11}}=-\omega^{\mu_10...\mu_{11}}=...
=\epsilon^{\mu_1\mu_2...\mu_{10}0}.\ff

\section*{Appendix B: Gamma matrix }
Some important features of Gamma matrices in eleven dimensional space
time are derived in this part. They are essential to show the closure of the
constraint algebra in present paper.
It's well known that the $\Gamma$-matrix plays an important role to
describe the spinor fields in various dimensions. They form the
Clifford algebra,
\f \Gamma^a\Gamma^c+\Gamma^c\Gamma^a=2\eta^{ac},\ff
we also introduce the notation
\f \Gamma^a\Gamma^c-\Gamma^c\Gamma^a:=2\Gamma^{ac},\ff
Then in the case of eleven dimensional space time, we can derive
the following identities
\f\Gamma_a\Gamma^a=11\label{GP1},\ff
\f\Gamma^a\Gamma^c=\Gamma^{ac}+\eta^{ac}\label{GP2}.\ff
($\ref{GP1}$) and ($\ref{GP2}$) are very useful when we try to
simplify the expression or rearrange the $\Gamma$-matrices into a
new order. For instance,
\f\Gamma_a\Gamma^d\Gamma^a=\Gamma_a(2\eta^{da}-\Gamma^a\G^d)=2\G^d-\G_a\G^a\G^d=-9\G^d,\ff
\f\G_a\G^{d_1...d_n}\G^a=(-1)^n(11-2n)\G^{d_1...d_n}.\ff
In this paper we also often use the following formula which are given
in \cite{Frebook},
\f\G^{a_1...a_nb}=\G^{a_1...a_n}\G^b-n\G^{[a_1...a_{n-1}}\eta^{a_n]b},\ff
\f\G^{ba_1...a_n}=\G^b\G^{a_1...a_n}-n\eta^{b[a_1}\G^{a_2...a_n]}.\ff

Next we give two important identities of Gamma matrices which are essential to
show the closure of constraint algebra. Both of them involve the exchanging of
spinor indices of different $\Gamma$-matrices, therefore we write down the elements of
these matrices labeled by the spinor indices explicitly. These identities are\footnote{We ignore an important matrix in all the paper, namely,
the charge conjugation matrix for a neat version of Gamma matrix.
But we'd better keep it in mind and realize it appeared where it
should be. In eleven dimensions it is also important to know that
only $\{\G^a, \G^{ab},\G^{a_1...a_5}\}$ and their dual matrices
are symmetric (under the action of charge conjugation matrix) while the others
are antisymmetric. Since in the
Poisson bracket of supersymmetry constraints, the charge
conjugation matrix is involved and any term which is
anti-symmetric will vanish. Therefore in the following equations we only write down
the symmetric term explicitly.}
\f {\G_a}^{(A}_{E}{\G^{ab}}^{B)}_{F}=-{1\over
4}(\G_a^{AB}\G^{ab}_{EF}+\G^{abAB}\G_{aEF}),\label{ident1}\ff
\f{\G^a}^{(A}_{E}{\G_{ab_1...b_4}}^{B)}_{F}-3{\G_{[b_1b_2}}^{(A}_{E}{\G_{b_3b_4]}}^{B)}_{F}
=-{1\over
4}(\G^{aAB}\G^{ab_1...b_4}_{EF}+\G_{ab_1...b_4}^{AB}\G^a_{EF})+{3\over
2}{\G_{[b_1b_2}}^{AB}\G_{b_3b_4]EF}, \label{ident2}\ff
It's not difficult to see, they go back to the ordinary Fiertz identities respectively
when four spinor fields are involved\cite{dAF},
\f\Psi_A\G^{aA}_B\Psi^B\Psi_E{\G_{ab}}^E_F\Psi^F=0\label{FI1},\ff
\f\Psi_A\G^{aA}_B\Psi^B\Psi_E{\G_{ab_1..b_4}}^E_F\Psi^F=3
\Psi_A\G^{A}_{[b_1b_2B}\Psi^B\Psi_E{\G_{b_3b_4]}}^E_F\Psi^F\label{FI2}.\ff

To prove identities $(\ref{ident1})$ and $(\ref{ident2})$, we need
apply the Fiertz decomposition formula in eleven dimensional space
time.
\f \G^{a(A}_E{\G_{ab}}^{B)}_F={1\over 32}\left[\G_d^{AB}(\G_a\G^d\G^{ab})_{EF}-{1\over
2}\G_{de}^{AB}(\G_a\G^{de}\G^{ab})_{EF}+{1\over
5!}\G_{d_1...d_5}^{AB}(\G_a\G^{d_1...d_5}\G^{ab})_{EF}\right].\label{decom1}\ff

Then our task is just to simplify the terms involving the
multiplication of several Gamma matrices in $(\ref{decom1})$.
Exploiting the formula given above, it's straightforward to derive the
following results,
\f \G_a\G^d\G^{ab}=-8\G^{db}+(anti-symmetric\ \ terms...),\label{d1}\ff
\f \G_a\G^{de}\G^{ab}=16\G^{[d}\eta^{e]b}+(anti-symmetric\ \ terms...),\label{d2}\ff
\f \G_a\G^{d_1...d_5}\G^{ab}=0+(anti-symmetric\ \ terms...).\label{d3}\ff
Substituting $(\ref{d1})$-$(\ref{d3})$ into $(\ref{decom1})$, we
easily arrive at the identity $(\ref{ident1})$.

As far as the second identity $(\ref{ident2})$ is concerned, we
just need do more complicated but straightforward calculations as
in the case of first identity.
\f \G^{a(A}_E{\G_{ab_1...b_4}}^{B)}_F={1\over 32}\left[\G_d^{AB}(\G_a\G^d\G^{ab_1...b_4})_{EF}-{1\over
2}\G_{de}^{AB}(\G_a\G^{de}\G^{ab_1...b_4})_{EF}+{1\over
5!}\G_{d_1...d_5}^{AB}(\G_a\G^{d_1...d_5}\G^{ab_1...b_4})_{EF}\right],\label{decom2}\ff
and
\f {\G_{[b_1b_2}}^{(A}_E{\G_{b_3b_4]}}^{B)}_F={1\over 32}\left[\G_d^{AB}(\G_{[b_1b_2}
\G^d\G_{b_3b_4]})_{EF}-{1\over 2}\G_{de}^{AB}(\G_{[b_1b_2}\G^{de}\G_{b_3b_4]})_{EF}+{1\over
5!}\G_{d_1...d_5}^{AB}(\G_{[b_1b_2}\G^{d_1...d_5}\G_{b_3b_4]})_{EF}\right].\label{decom3}\ff
In $(\ref{decom2})$, the terms in brackets can be simplified
respectively as,
\f \G_a\G^d\G^{ab_1...b_4}=-5\G^{db_1...b_4}+anti-sym.\ \ terms,\ff
\f \G_a\G^{de}\G^{ab_1...b_4}=3\G^{deb_1...b_4}+84\eta^{e[b_1}\eta^{|d|b_2}
\G^{b_3b_4]}+anti-sym.\ \ terms,\ff
\fa \G_a\G^{d_1...d_5}\G^{ab_1...b_4}&=&3\G^{d_1...d_5}\G^{b_1...b_4}-40\G^{[d_1...d_4}
\eta^{d_5][b_1}\G^{b_2...b_4]}\nonumber\\&=&3\G^{d_1...d_5}_{b_1...b_4}
-5!\delta^{[d_1}_{[b_2}\delta^{d_2}_{b_1}\G^{d_3d_4d_5]}_{b_3b_4]}+5\cdot5!
\G^{[d_1}\delta^{d_2}_{[b_1}\delta^{d_3}_{b_2}\delta^{d_4}_{b_3}\delta^{d_5]}_{b_4]}
\nonumber\\&&+anti-sym.\ \
 terms,\ffa
and in (\ref{decom3}), the terms in brackets can be simplified as
\f\G_{[b_1b_2}\G^d\G_{b_3b_4]}=\G^d_{b_1...b_4}+anti-sym.\ \ terms,\ff
\f\G_{[b_1b_2}\G^{de}\G_{b_3b_4]}=\G^{de}_{b_1...b_4}-4 \delta^{[d}_{[b_2}\delta^{e]}_{b_1}
\G_{b_3b_4]}+anti-sym.\ \ terms,\ff
\f\G_{[b_1b_2}\G^{d_1...d_5}\G_{b_3b_4]}=\G^{d_1...d_5}_{b_1...b_4}
-40\delta^{[d_1}_{[b_2}\delta^{d_2}_{b_1}\G^{d_3d_4d_5]}_{b_3b_4]}+5!\delta^{[d_1}_{[b_2}
\delta{d_2}_{b_1}\G^{d_3}\delta^{d_4}_{b_4}\delta^{d_5]}_{b_3]}.\ff

Substituting all the terms into (\ref{decom2}) and (\ref{decom3})
respectively, we will find the identity (\ref{ident2}) holds
indeed.

\section*{Appendix C: The proof of the closure of constraint algebra}
In this section we only show the closure of two Poisson brackets.
One involves the Gaussian constraint, and the other is the
supersymmetric constraint. The other Poisson brackets are closed
trivially.
First we consider the Poisson bracket of Gaussian constraint. To
make the calculation clear, we divide the constraint into two
parts,
 \f G^a =:G^a_1+G^a_2,\ff
Where
\f G^a_1= D_k\pi^{ka},\ff
and
\f G^a_2=-\Psi^{A}_{i}\Psi^{B}_{j} \left \{ \Gamma^{a}_{bAB} (
R^{ijk}+{15\over 3!} p^{ijkl_{3}}a_{l_{3}} ) e_{k}^{b} + {5 \imath
\over 2} \Gamma^{a}_{b_{4}AB}p^{ijk_{4}}E_{k_{4}}^{b_{4}} \right
\}.\ff
It's straightforward to compute the Poisson brackets of them,
\f \{G^a_1, G^b_2\}=0,\label{G1}\ff
\f\{G^a_2,G^b_2\}=\delta^{10}30\Psi_{iA}\G^{acA}_{B}\Psi^B_j\Psi_{mE}\G^{bdE}_D\Psi^D_n
e_{kc}e_{pd}p^{mnpijk},\label{c1}\ff
\fa\{G^a_1,G^b_2\}&+&\{G^a_2,G^b_1\}=\delta^{10}\left\{ \left [15\Psi_{iA}\G^{abA}_{B}\Psi^B_j\Psi_{mE}\G^{cdE}_D\Psi^D_n
e_{kc}e_{pd}p^{mnpijk}\right.\right.\label{c2}\\&&-\left.\left.15\Psi_{iA}\G^{abcdeA}_{B}\Psi^B_j\Psi_{kC}\G^C_{cD}\Psi^D_p
E_{demn}p^{ijkmnp}\right ]\right.\label{c3}\\&&+\left.
\Gamma^{ab}_{AB}\left [2F_{ki}^{A}\Psi_{j}^{B}(R^{ijk} + {15\over 3!}
p^{ijkl_{3}}a_{l_{3}}) + \Psi^{A}_{i}\Psi^{B}_{j}( {1\over
4}G_{2}^{ij}+G_{5}^{ijl_{3}}a_{l_{3}}
+p^{ijl_{4}}F_{l_{4}}^{\Box}) \right ]\right. \nonumber\\
&&+\left.\Gamma^{ab}_{defAB}\left[ {2\imath }F_{ki}^{A}\Psi_{j}^{B}
p^{ijkl_{3}}E_{l_{3}}^{def}+{\imath \over 4}
G_{5}^{ijl_{3}}E_{l_{3}}^{def}+{\imath }
p^{ijklm_{2}}F_{kl^{d}}E_{m_{2}}^{ef} \right ] \right\}.\label{c4}\ffa
Now add $(\ref{c1})$ and $(\ref{c2})$ together and notice that
\f \G^{[ab}\G^{cd]}={1\over
3}(\G^{ab}\G^{cd}+\G^{ac}\G^{db}+\G^{ad}\G^{bc}).\ff
we find the sum of three terms $(\ref{c1})$, $(\ref{c2})$ and
$(\ref{c3})$ vanishes by employing the standard Fiertz identity
$(\ref{FI2})$. Making a collection of $(\ref{G1})$-$(\ref{c4})$,
we show the closure of Poisson bracket of Gauss
constraint which corresponds to (\ref{GwG}) in the paper.

The Poisson bracket $(\ref{SPB})$ can be derived in a similar way
except that we need apply the identities (\ref{ident1}) and
(\ref{ident2}) to cancel those extra terms.

The supersymmetric constraint is
\f Q^A=Q_1^A+Q_2^A,\ff
where
\f Q_1^A=D_i\pi^{iA} -{\imath }\pi^{ia}\Gamma^{A}_{aB}\Psi^B_{i},\ff
and
\f Q_2^A= -\Psi^{B}_{i}\left \{
 \Gamma_{abB}^A ( R^{ijk}+{15\over 3!} p^{ijkl_{3}}a_{l_{3}}
)E^{ab}_{jk} -{\imath } \Gamma^A_{a_{5}B}
p^{ik_{5}}E_{k_{5}}^{b_{5}} \right \}.\ff
We find the Poisson brackets of them are
\f\{Q_1^A, Q_1^B\}=i\delta^{10}\G_a^{AB}D_i\pi^{ia},\label{I}\ff
\f\{Q_2^A,
Q_2^B\}=30\delta^{10}\Psi_i^C\Psi_j^D\G_{[abC}^{(A}\G_{cd]D}^{B)}E^{abcd}_{kmnp}p^{ijkmnp},\label{I2}\ff
\fa\{Q_1^A, Q_2^B\}&+&\{Q_2^A+Q_1^B\}=\delta^{10}\left\{
D_i[\G^{AB}_{ab}e^{ab}_{jk}( R^{ijk}+{15\over 3!} p^{ijkl_{3}}a_{l_{3}}
)+i\G^{AB}_{a_5}E^{a_5}_{j_5}p^{ij^5}]\right.\nonumber\\&&+\left.4i\Psi^C_i\Psi^D_j(\G_{aC}^{(A}\G^{B)ab}_{D}e_{bk}
( R^{ijk}+{15\over 3!} p^{ijkl_{3}}a_{l_{3}})+{5i\over
2}\G^{(A}_{aC}\G^{B)ab_4}E^{b_4k_4}p^{ijk_4})\right\}.\label{S1}\ffa
In (\ref{S1}), it's a little tedious to deal with the terms with
covariant derivative. To express them as the sum of constraints,
we use the fact that
\f D_{[\mu}e^a_{\nu]}={1\over 2}F^a_{\mu\nu}+{i\over
2}\Psi_{A[\mu}\G^{aA}_B\Psi^B_{\nu]},\ff
and
\f \partial_{[\mu}a_{\nu\rho\sigma]}={1\over
4}F^{\Box}+{3!\over2}\Psi_{A[\mu}\G^{abA}_{B}\Psi^B_{\nu}e_{|a|\rho}e_{|b|\sigma]},\ff
and then expand those terms as follows,
\fa &&D_i[\G^{AB}_{ab}e^ab_{jk}(R^{ijk}+{15\over
3!}a_{mnp}p^{ijkmnp})]=(...)F^a+(...)G^{ij}+(...)G^{i_5}+(...)F^{\Box}_{imnp}
\nonumber\\&&+i\G^{AB}_{ab}\Psi_{ic}\G^{aC}_D\Psi^D_je^b_k(R^{ijk}+{15\over
3!}a_{mnp}p^{ijkmnp})+15\G^{AB}_{ab}e^{ab}_{jk}\Psi_{iC}\G^C_{cdD}
\Psi^D_me^{cd}_{np}p^{ijkmnp},\ffa
\f
D_i(\G^{AB}_{a_5}E^{a_5}_{j_5}p^{ij^5})=(...)G^{j_5}+(...)F^a+{5i\over
2}\G^{AB}_{ab_4}\Psi_{iC}\G^{aC}_D\Psi^D_jE^{b_4}_{k_4}p^{ijk_4},\ff
where we ignore the explicit expressions of terms involving
curvatures and constraints, but the final results are given
in (\ref{SPB}).
Making use of identities $(\ref{ident1})$, we add the terms
containing $( R^{ijk}+{15\over 3!} p^{ijkl_{3}}a_{l_{3}}
)$ together in the brackets of $(\ref{S1})$ and have
\f i\Psi_i^C\Psi_j^De_{bk}( R^{ijk}+{15\over 3!} p^{ijkl_{3}}a_{l_{3}}
)(4\G^{(A}_{aC}\G^{B)ab}_D+\G^{AB}_{ab}\G^a_{CD})=-i\Psi_i^C\Psi_j^De_{bk}( R^{ijk}+{15\over 3!} p^{ijkl_{3}}a_{l_{3}}
)\G^{AB}_a\G^{ab}_{CD},\label{II}\ff
and using $(\ref{ident2})$, we pick out all the terms containing
$p^{ijkmnp}$ in the algebra $(\ref{I2})$ and $(\ref{S1})$, and find
\fa
&&\Psi^C_i\Psi^D_jE^{abcd}_{kmnp}p^{ijkmnp}(30\G^{(A}_{C[ab}\G^{B)}_{cd]D}-10
\G^{(A}_{Ce}\G^{B)eabcd}_D-{5\over
2}\G^{AB}_{eabcd}\G^e_{CD}+15\G^{AB}_{[ab}\G_{cd]CD})\nonumber\\&&
={5\over
2}\Psi^C_i\Psi^D_jE^{b_4}_{k_4}p^{ijkmnp}\G^{AB}_a\G^a_{b_4CD}.\label{III}\ffa

Next combining all the terms in (\ref{I}), (\ref{II}) and (\ref{III})
together, we find it's nothing but the Gauss constraint $G^a$!
After collecting the other terms remaining which contain curvatures and constraints
in the algebra we finally arrive at $(\ref{SPB})$, which is what we need to show.

\end{document}